\documentclass{appolb}

\usepackage{multirow}
\usepackage{graphicx}
\usepackage{amsmath,amssymb,color,hyperref}

\def\ttt#1{\texttt{\scriptsize #1}}

\providecommand{\epem}{e^{+}e^{-}}
\providecommand{\pp}{p-p}

\providecommand{\pPb}{p-Pb}

\providecommand{\PbPb}{Pb-Pb}
\providecommand{\gaga}{\gamma\,\gamma}

\providecommand{\mgg}{\rm m_{\gamma\gamma}}
\providecommand{\sigmagg}{\sigma_{\gaga\to\gaga}}

\newcommand{\sqrtsgg}{\sqrt{s_{_{\gamma\,\gamma }}}}
\newcommand{\sqrtsnn}{\sqrt{s_{_{\rm NN}}}}

\newcommand{\ABgaga}{A\,B\,$\xrightarrow{\gaga}$ A\,$\gaga$\,B}
\newcommand{\ABglgl}{A\,B\,$\xrightarrow{g\,g}$ A\,$\gaga$\,B}

\providecommand{\madgraph}{{\sc MadGraph}}
\providecommand{\superchic}{{\sc SuperChic}}

\newcommand{\Lumi}{\mathcal{L}}

\newcommand{\Pom} {I\!P} 

\begin{document}


\title{Measuring light-by-light scattering at the LHC and FCC\thanks{Proceedings EDS'15 Blois Conference,
Corsica, June 2015}}
\author{David~d'Enterria
\address{\vspace{-0.35cm}CERN, EP Department, 1211 Geneva, Switzerland}\\\vspace{0.25cm}
{Gustavo G. da~Silveira}
\address{\vspace{-0.35cm}Instituto de F\'isica e Matem\'atica, Univ. Fed. de Pelotas, Caixa Postal 354, \\CEP 96010-090, Pelotas, RS, Brazil}
}
\maketitle
\begin{abstract}
Elastic light-by-light scattering, $\gaga\to\gaga$, can be measured 
in electromagnetic interactions of lead (Pb) ions at the Large Hadron Collider (LHC) and Future Circular
Collider (FCC), using the large (quasi)real photon fluxes available in ultraperipheral collisions. The $\gaga~\to~\gaga$
cross sections for diphoton masses $\mgg>$~5~GeV in \pp, \pPb, and \PbPb\ collisions at LHC ($\sqrtsnn$~=~5.5,
8.8, 14~TeV) and FCC ($\sqrtsnn$~=~39, 63, 100~TeV) center-of-mass energies are presented. 
The measurement has controllable backgrounds in \PbPb\ collisions, and one expects about 70 and 2\,500
signal events per year at the LHC and FCC respectively, after typical detector acceptance and reconstruction
efficiency selections.
\end{abstract}


\section{Introduction}

\begin{figure}[hbt!]
\centering
\includegraphics[height=4.5cm]{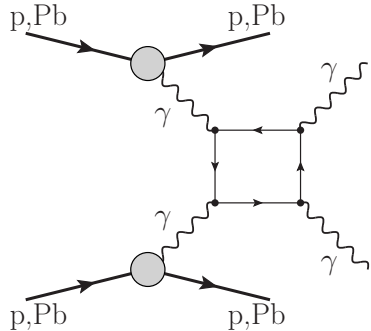}
\caption{Diagram of elastic $\gaga\to\gaga$ collisions in ultraperipheral proton and/or ion interactions.
 The initial-state photons are emitted coherently by the protons and/or nuclei, which survive the electromagnetic interaction.} 
  \label{fig:diag}
\end{figure}

The elastic scattering of two photons in vacuum, $\gaga\to\gaga$, is a pure quantum mechanics process that
proceeds at leading order (LO) in the fine structure constant, $\mathcal{O}(\alpha^4)$,
via virtual box diagrams containing charged particles. In the standard model (SM), the box diagram of
Fig.~\ref{fig:diag} involves charged fermions (leptons and quarks) and boson (W$^\pm$) loops. 
Despite its simplicity, light-by-light (LbyL) scattering remains still unobserved today because of its tiny
cross section $\sigma_{\gaga}\propto\mathcal{O}(\alpha^4)\approx$~3$\cdot 10^{-9}$, although the electron
loop contribution has been precisely, yet indirectly, tested in the anomalous electron and muon magnetic
moments measurements. Out of the two closely-related processes --photon scattering in the Coulomb field of a 
nucleus (Delbr\"uck scattering) and photon-splitting in a strong magnetic field (vacuum birefringence)--
only the former has been experimentally observed~\cite{Jarlskog:1974tx}. 
Apart from the intrinsic importance of its direct observation in the laboratory, $\gaga$ scattering provides a
particularly neat channel to study anomalous gauge couplings~\cite{Brodsky:1994nf}, and search for 
physics beyond the SM (BSM) through new heavy charged particles contributing to the virtual loop in Fig.~\ref{fig:diag}
--in particular at high diphoton invariant masses--  
such as \eg SUSY particles~\cite{Gounaris:1999gh}. LbyL scattering has also been proposed as a means to 
search for monopoles~\cite{Ginzburg:1998vb}, axions~\cite{Bernard:1997kj}, unparticles~\cite{Kikuchi:2008pr},
low-scale gravity effects~\cite{Cheung:1999ja}, and non-commutative
interactions~\cite{Hewett:2000zp}.

Several approaches have been proposed to
experimentally detect $\gaga\to\gaga$ using \eg Compton-backscattered photons against laser
photons~\cite{Mikaelian:1981fh}, collisions of photons from  microwave
waveguides/cavities~\cite{Brodin:2001zz} or high-power lasers~\cite{Lundstrom:2005za}, as well as at photon 
colliders~\cite{Brodsky:1994nf} where energetic $\gamma$ beams can be obtained by Compton-backscattering
laser-light off $\epem$ beams. In~\cite{d'Enterria:2013yra} we demonstrated that one can detect
elastic $\gaga$ scattering using the large (quasi)real photon fluxes of the protons and ions
accelerated at TeV energies at the CERN Large Hadron Collider (LHC). In this work, we summarize the
results of~\cite{d'Enterria:2013yra} and extend the study for the energies of the Future Circular Collider
(FCC), a new facility proposed for BSM searches in a new 80--100~km tunnel to be
constructed at CERN~\cite{fcchh}.

\section{Theoretical setup}

All charges accelerated at high energies generate electromagnetic fields which, in the equivalent photon
approximation (EPA)~\cite{WW}, can be considered as $\gamma$ beams of virtuality $- Q^{2} < 1/R^{2}$, where
$R$ is the radius of the charge, \ie $Q^2\approx$~0.08~GeV$^2$ for protons ($R\approx$~0.7~fm), and
$Q^2<$~4$\cdot10^{-3}$~GeV$^2$ for nuclei ($R_{\rm A}\approx 1.2\,A^{1/3}$~fm, for mass number
$A>$~16). The photon spectra have a typical $E_{\gamma}^{-1}$ power-law fall-off up to energies of the order of
$\omega_{\rm max}\approx\gamma/R$, where $\gamma$ is the Lorentz relativistic factor of the proton 
or ion. 
Although the $\gamma$ spectrum is harder for smaller charges --which favours proton over nuclear beams in the
production of heavy diphoton systems-- each photon flux scales with the squared charge of the beam, $Z^2$, and
thus $\gaga$ luminosities are extremely enhanced, up to $Z^4$~=~5$\cdot$10$^{7}$ in the case of
\PbPb, for ion beams. Table~\ref{tab:1} summarizes the relevant parameters for ultraperipheral \pp, \pPb,
and \PbPb\  collisions at the LHC and FCC. Two-photon center-of-mass (c.m.) energies at the FCC will reach for the
first time the multi-TeV range.\\
\begin{table}[htpb]
\begin{center}
\caption[]{Characteristics of $\gaga\to\gaga$ measurements in A\,B collisions at LHC and FCC:
(i) nucleon-nucleon c.m.{} energy, $\sqrtsnn$, (ii) integrated luminosity $\Lumi_{\rm AB}\cdot\Delta t$ 
($\Lumi_{\rm AB}$ are beam luminosities --for low pileup in the \pp\ case-- and a ``year''
is $\Delta t$~=~10$^{7}$~s for \pp, and 10$^{6}$~s in the ion mode), 
(iii) beam Lorentz factor, $\gamma$, 
(iv) maximum photon energy in the c.m.{} frame, $\omega_{\rm max}$, 
(v) maximum photon-photon c.m.{} energy, $\sqrt{s_{\gaga}^{\rm max}}$, (vi) 
cross section for $\gaga$ masses above 5~GeV, and (vii) expected number of counts/year after selection cuts.} 
\label{tab:1}
\begin{tabular}{l|ccccc|cc} \hline 
\hspace{-0.4cm} System \hspace{-0.2cm} & \hspace{-0.1cm} $\sqrtsnn$ \hspace{0.01cm} & \hspace{0.01cm} $\mathcal{L}_{\rm AB}\cdot\Delta t$ \hspace{0.01cm} 
& \hspace{0.01cm} $\gamma$ \hspace{0.01cm} 
& \hspace{0.01cm} $\omega_{\rm max}$ \hspace{0.01cm} 
& \hspace{0.01cm} $\sqrt{s_{\gaga}^{\rm max}}$ \hspace{0.01cm} 
& \hspace{0.01cm} $\sigmagg^{\rm excl}$\hspace{0.01cm} 
& \hspace{0.01cm} $N_{\gaga}^{^{\rm cuts}}$ \hspace{0.01cm} \\
       & (TeV) &  (per year) &  ($\times$10$^3$) & (TeV) & (TeV) & \multicolumn{2}{c}{[$\mgg>$5~GeV]} \\ \hline
\hspace{-0.3cm} \pp    & 14  & 1~fb$^{-1}$  & 7.5  & 2.45 & 4.5  & 105 $\pm$ 10 fb &  12 \\ 
\hspace{-0.3cm} \pPb   & 8.8 & 200~nb$^{-1}$& 4.7  & 0.13 & 0.26 & 260 $\pm$ 26 pb &   6 \\ 
\hspace{-0.3cm} \PbPb  & 5.5 & 1~nb$^{-1}$  & 2.9  & 0.80 & 0.16 & 370 $\pm$ 70 nb &  70 \\ \hline
\hspace{-0.3cm} \pp    & 100 & 1~fb$^{-1}$  & 53.  & 17.6 & 35.2 & 240 $\pm$ 24 fb &  50 \\ 
\hspace{-0.3cm} \pPb   &  63 & 1~pb$^{-1}$  & 33.5 & 0.95 & 1.9  & 780 $\pm$ 78 pb & 150 \\ 
\hspace{-0.3cm} \PbPb  &  39 & 5~nb$^{-1}$  & 21.  & 0.60 & 1.2  & 1.85$\pm$ 0.37 $\mu$b & 2\,500\\ \hline
\end{tabular}
\end{center}
\end{table}

Photon-photon collisions in ``ultraperipheral'' collisions (UPCs) of proton~\cite{d'Enterria:2008sh} 
and lead (Pb) beams~\cite{Baltz:2007kq} have been experimentally measured at the
LHC~\cite{Chatrchyan:2012tv,Abbas:2013oua,Aad:2015bwa}.
The UPC final-state signature in this work is the exclusive production of two photons, \ABgaga, with
the diphoton final-state measured in the central detector, and the hadrons A,B~=~p,Pb surviving the
electromagnetic interaction scattered at very low angles with respect to the beam. The very same
final-state can be mediated by the strong interaction through a quark-loop in the exchange of two gluons in a
colour-singlet state, \ABglgl~\cite{superchic}. Such ``central exclusive production'' (CEP), observed in
p$\bar{\rm p}$ at Tevatron~\cite{Aaltonen:2011hi} and searched for at the
LHC~\cite{Chatrchyan:2012tv}, constitutes an important background for the $\gaga\to\gaga$ measurement in \pp\
but not \PbPb\ collisions as discussed later.
In the EPA, the elastic $\gaga$ production cross section in UPCs of
hadrons A and B factorizes into the product of the elementary $\gaga\rightarrow \gaga$ cross section at
$\sqrtsgg$, convolved with the photon fluxes $f_{\rm \gamma/A,B}(\omega)$ of the two colliding beams: 
\begin{equation}
\sigmagg^{\rm excl}=\sigma(\rm{A} \rm{B} \xrightarrow{\gaga} \rm{A} \gaga \rm{B})=
\int d\omega_1 d\omega_2  \frac{f_{\rm \gamma/A}(\omega_1)}{\omega_1} \frac{f_{\rm \gamma/B}(\omega_2)}{\omega_2} \sigmagg(\sqrtsgg),
\label{eq:two-photon}
\end{equation}
where $\omega_1$ and $\omega_2$ are the energies of the photons emitted by hadrons A and B. We use the proton 
f$_{\rm \gamma/p}(\omega)$ spectrum derived from its elastic form factor~\cite{Budnev:1974de} and, 
the impact-parameter dependent expression for the ion f$_{\rm \gamma/A}(\omega)$
spectrum~\cite{Bertulani:1987tz}, including a correction equivalent to ensuring that all collisions are purely
exclusive, \ie without hadronic overlap and breakup of the colliding beams~\cite{Cahn:1990jk}.
The \madgraph\ v.5 Monte Carlo (MC)~\cite{madgraph} framework is used to convolve the 
$\gamma$ fluxes as done in~\cite{d'Enterria:2009er}, with the LO expression for the $\sigmagg$ cross
section~\cite{Bern:2001dg} including all quark and lepton loops. We omit the W$^\pm$ contributions which are
only important at $\mgg\gtrsim$~200~GeV. 
Inclusion of next-to-leading-order QCD and QED corrections increases $\sigmagg$ by a few
percent~\cite{Bern:2001dg}, which 
are nonetheless ``compensated'' by the $\hat{S}^2$~=~0.9--1.0  gap survival factor
--encoding the probability to produce the $\gaga$ system without any other hadronic
activity from soft rescatterings between the colliding hadrons~\cite{superchic}-- that reduces the 
exclusive yields by about the same amount. Propagated uncertainties to the final cross sections are of order
$\pm$10\% ($\pm$20\%) for \pp\ and \pPb\ (\PbPb) collisions, covering different form-factors parametrizations
and the convolution of the nuclear photon fluxes.\\ 

Our calculations require diphoton masses in the continuum above $\mgg$~=~5~GeV, avoiding the region of
two-photon decays from CEP hadronic resonances (\eg $\chi_{\rm c0,c2}\to\gaga$ at masses 3.4--3.9~GeV), and
where one can easily define an experimental trigger based on a few GeV deposit in the calorimeters. Also, for
lower diphoton masses, the $\gaga$ cross section has larger theoretical uncertainties as the hadronic LbyL
contributions are computed less reliably by the quark boxes~\cite{Bern:2001dg}.

\section{Signal cross sections}

\begin{figure}[htbp]
\centering
\includegraphics[width=0.6\columnwidth,height=7.35cm]{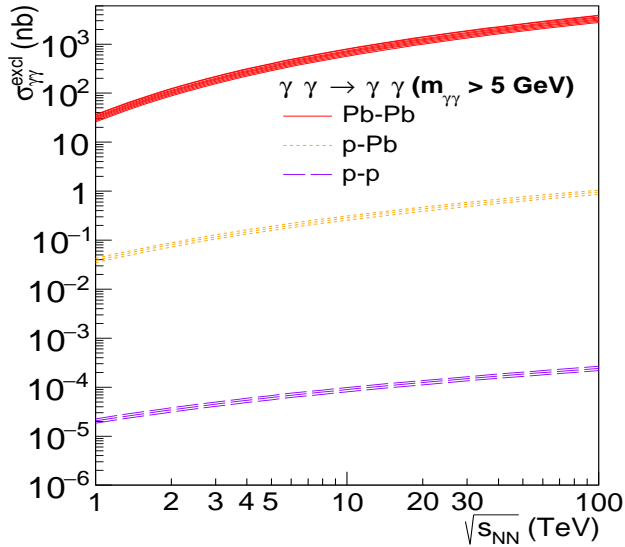}
\caption{Cross sections for $\gaga\to\gaga$, with pair masses above 5~GeV, in ultraperipheral \PbPb\
(top curve), \pPb\ (middle) and \pp\ (bottom) collisions as a function of c.m.{} energy.}
\label{fig:sigma}
\end{figure}

Using the theoretical setup described, we obtain the values of 
$\sigmagg^{\rm excl}$ 
listed in Table~\ref{tab:1} 
and plotted as a function of c.m.{} energies in the range $\sqrtsnn$~=~1--100~TeV in Fig.~\ref{fig:sigma}.
The cross sections are in the hundreds of fb/pb/nb for \pp, \pPb, and \PbPb\ (and even, remarkably, at the
$\mu$b level for the latter at the FCC), clearly showing the importance of the $Z^4$ photon-flux enhancement for ions
compared to protons. The increase in cross sections from LHC to FCC is of $\mathcal{O}$(2--5).
The detectable number of $\gaga\to\gaga$ events at the LHC and FCC are estimated by considering nominal
luminosities for each system, geometric detector acceptance, and reconstruction efficiencies.
For the LHC (FCC) we consider photon detection capabilities with tracking and calorimetry over
pseudorapidities $|\eta\,|<$~2.5~(5.0), plus forward detectors, up to at least $|\eta|$~=~5, to select
exclusive events requiring rapidity-gaps on both sides of the central diphoton system. The requirement to have
both photons with $\rm p_{T}^{\gamma}>$~2~GeV within the $|\eta|$ ranges considered, reduces the yields by
$\varepsilon_{\rm acc}\approx$~0.2 (0.3) in \pp\ and \pPb, and by $\varepsilon_{\rm acc}\approx$~0.3 (0.4) in
\PbPb\ collisions at the LHC (FCC). 
The acceptance is larger in the \PbPb\ case because the EPA fluxes are softer and the $\gaga$ 
system is produced at more central rapidities. We also consider offline $\gamma$ reconstruction and
identification efficiencies $\varepsilon_{\rm rec,id\,\gamma}\approx$~0.8 in the photon energy range
of interest. The final combined signal efficiency is
$\varepsilon_{\rm pp,pPb\to\gaga}=\varepsilon_{\rm trig}\cdot\varepsilon_{\rm acc}\cdot\varepsilon_{\rm
  rec,id\,\gamma}^2\approx$~12\% (20\%) for \pp\ and \pPb, and $\varepsilon_{\rm PbPb\to\gaga}\approx$~20\%
(26\%) for \PbPb, at the LHC (FCC). The number of events expected per year ($\Delta t$~=~10$^{6}$~s in the 
ion mode, 10$^{7}$~s for \pp) are obtained via 
$N_{\gaga}^{\rm excl}=\varepsilon_{\gaga}\cdot\sigma_{\gaga}^{\rm excl}\cdot\Lumi_{\rm AB}\cdot\Delta t$ (Table~\ref{tab:1}). 
The nominal \pPb\ and \PbPb\ luminosities are low enough to keep the number of simultaneous collisions 
well below one, and one can take their full integrated luminosity 
as usable for the exclusive measurement. In \pp, pileup is very  high and we 
consider that only 1~fb$^{-1}$/year can be collected under conditions 
that preserve the rapidity gaps 
adjacent to the central $\gaga$ system.
Clearly, \PbPb\ provides the best signal counting rates, with statistical uncertainties 
of order $\sqrt{\rm N_{\gaga}^{\rm excl}} = \pm$12\%$_{\rm (LHC)}$,2\%$_{\rm (FCC)}$, free of pileup complications.

\section{Backgrounds, and $\gaga\to\gaga$ significances}

There are three potential backgrounds that share the same (or very similar) final-state
signature as $\gaga\to\gaga$: (i) central-exclusive diphoton production\footnote{For CEP $\pi^0\pi^0$ and
  $\eta^{(')}\eta^{(')}$, decaying into multi-photon final-states,
their $\gamma$ branching ratios, acceptance plus $\mgg$
cuts results in a negligible final contribution compared to CEP $\gaga$~\cite{Harland-Lang:2013ncy}.}
$gg\to\gaga$, (ii) QED $\gaga\to\epem$ events, with both $e^{\pm}$ misidentified as photons, and
(iii) diffractive Pomeron-induced ($\Pom\Pom$, or $\gamma\Pom$) processes with final-states containing two
photons plus rapidity gaps. The latter diffractive and $\gamma$-induced final-states have larger 
p$_{\rm T}^{\gaga}$ and diphoton acoplanarities than $\gaga\to\gaga$, and can be efficiently removed.
The CEP $gg\to\gaga$ background, however, scales with the fourth power of the gluon density and is a large
potential background. In the \pp\ case and for the range of $\mgg$ considered here, we obtain a cross section
after acceptance cuts of $\sigma_{gg\to\gaga}^{\rm CEP}$~=~20$^{\times 3}_{\times 1/3}$~pb at the LHC with
\superchic~1.41~\cite{superchic}, where the large uncertainties include the choice of the parton
distribution function (PDF) and $\hat{S}^2$ survival factor. Typical CEP photon pairs peak at 
p$_{\rm T}^{\gaga}\approx$~0.5~GeV and have moderate tails in their azimuthal acoplanarity
$\Delta\phi_{\gaga}$, whereas photon-fusion systems are produced almost at rest. 
By imposing very tight cuts in the pair momentum, p$_{\rm T}^{\gaga}\lesssim$~0.1~GeV and acoplanarity
$\Delta\phi_{\gaga}-\pi\lesssim$~0.04, the CEP $\gaga$ can be reduced by a factor of about $\times$90 
with minimum losses 
of the elastic $\gaga$ signal. However, the resulting LbyL/CEP~$\approx$~1/20
ratio is still too large to make feasible the LbyL observation with proton beams. The situation is more advantageous
for \pPb, where the LbyL cross section is only about 6 times smaller than the CEP one, obtained scaling by
A~=~208 the \pp\ cross section at 8.8~TeV ($\sigma_{gg\to\gaga}^{^{\rm CEP}}$~=~16$^{\times 3}_{\times 1/3}$~pb),
multiplied by the square of the Pb gluon shadowing, $R_g^{^{\rm Pb/p}}\approx$~0.7 according to the
EPS09 nuclear PDF~\cite{eps09}. A final LbyL/CEP~$\approx$~1 ratio is reachable applying the aforementioned
p$_{\rm T}^{\gaga}$ and $\Delta\phi_{\gaga}$ cuts. Yet, given the low \pPb\ rates expected
at the LHC (Table~\ref{tab:1}), a 5-$\sigma$ observation of LbyL scattering requires an increase
of the luminosity from its conservative nominal value~\cite{d'Enterria:2009er}.\\

In the \PbPb\ case, the situation is more favourable given that parton-mediated 
exclusive or diffractive cross sections (which scale as A$^2$ compared to \pp) play a
comparatively smaller role than in \pp\ thanks to the Z$^{4}$-enhancement of electromagnetic
interactions. The \PbPb\ CEP cross section, as obtained by  A$^2$-scaling the 
$\sigma_{gg\to\gaga}^{\rm CEP}$~=~13$^{\times 2.5}_{\times 0.4}$~pb cross section in \pp\ at 5.5~TeV,
multiplied by $(R_g^{^{\rm Pb/p}})^{4}\approx$~0.25, is comparable to $\sigmagg^{\rm excl}$. Adding a simple
p$_{\rm T}^{\gaga}<$~0.2~GeV condition, reduces the CEP background by $\sim$95\% without removing any signal
event, resulting in a final  LbyL/CEP~$\approx$~10 ratio.
Other electromagnetic processes in \PbPb\ are, however, similarly enhanced by the Z$^4$ factor
and can constitute a potential background if the final-state particles are misidentified as photons.
The very large exclusive \PbPb$\xrightarrow{\gaga}\epem$ QED cross section, 
$\sigma_{\rm \gaga\to \epem}^{\rm QED}$[m$_{\rm \epem}>$~5~GeV]~=~5.4~mb according to
\textsc{Starlight}~\cite{Nystrand:2004vn}, can be of concern if neither $e^\pm$ track is reconstructed or if
both $e^\pm$ undergo hard bremsstrahlung. 
Requiring both $e^\pm$ to fall within the central acceptance and be singly misidentified as photons with
probability $f_{\rm e\to\gamma}\approx$~0.5\%, 
results in a residual $\gaga\to\gamma_{(e^+)}\,\gamma_{(e^-)}$ contamination of $\sim$20\% of the visible
LbyL cross section. In \PbPb\ at FCC(39~TeV), the CEP cross section within $|\eta|<$~5 is very large:
$\sigma_{gg\to\gaga}^{\rm CEP}$[m$_{\gaga}>$\,5~GeV] = $1.3$~nb~$\times 208^2 \times (R_g^{^{\rm Pb/p}})^4$
$\approx$~14~$\mu$b (with a factor of $\sim$3 uncertainty) obtained with \superchic~2.02~\cite{superchic2} and the
MMHT2014 PDFs~\cite{MMHT2014}, reduced to $\sim$400~nb after cuts. 
The QED cross section 
is $\sigma_{\rm \gaga\to \epem}^{\rm QED}$[m$_{\rm \epem}>$\,5~GeV] = 26~mb, reduced to $\sim$120~nb after
applying the $f_{\rm e\to\gamma}^2$ factor and acceptance selection criteria. After cuts, both backgrounds
are thereby smaller than the expected visible LbyL cross section of $\sim$500~nb.

\begin{figure}[htbp]
\centering
\includegraphics[width=0.495\columnwidth]{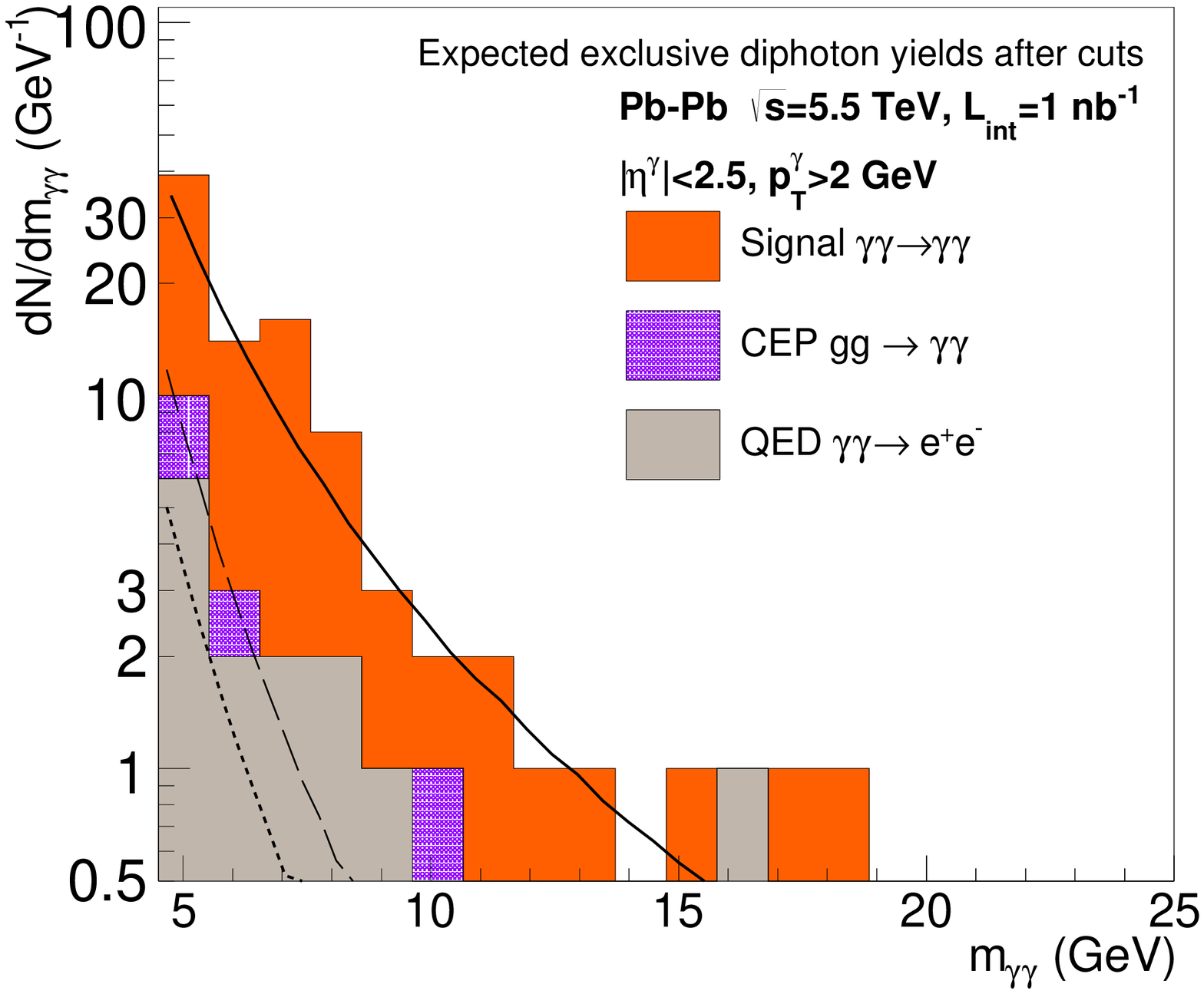}
\includegraphics[width=0.495\columnwidth]{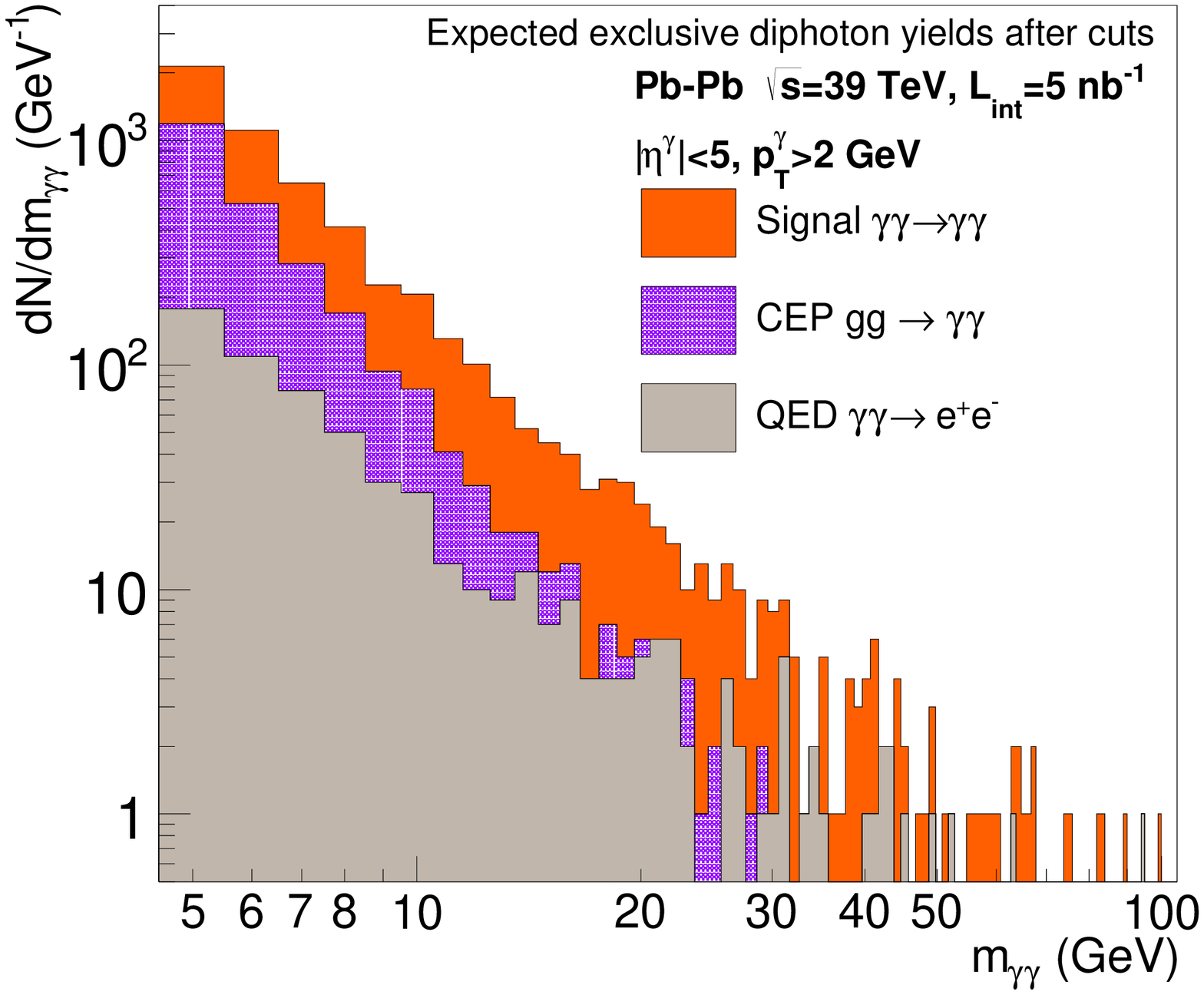}
\caption{Yields as a function of diphoton invariant mass for elastic $\gaga$, plus CEP-$\gaga$ and
QED backgrounds, expected in \PbPb\ at LHC (left) and FCC (right)
after analysis cuts.}
\label{fig:minv}
\end{figure}

Figure~\ref{fig:minv} shows the $\gaga$ invariant mass distributions for signal
and CEP and QED backgrounds after cuts in one \PbPb\ run at the LHC (left)~\cite{d'Enterria:2013yra} and FCC (right). 
At the LHC (FCC), we expect about $\rm N_{\gaga}^{\rm excl}\approx$~70 (2\,500) signal counts compared to
$\sim$6 (2\,000) and $\sim$15 (600) CEP and QED counts respectively, with FCC reaching much higher diphoton masses.
The overall (profile likelihood) significances of the measurement are ${\cal S}\approx$~6 at the LHC and
${\cal S}\approx$~35 at FCC, considering 20\% and 50\% theoretical uncertainties on LbyL and CEP yields
respectively (the QED $\epem$ background can be easily well measured beforehand).

\section{Summary}

We have shown that light-by-light scattering, a rare fundamental 
process that has escaped experimental observation so far, can be measured at the LHC and FCC exploiting the
large quasireal photon fluxes in electromagnetic interactions of protons and ions accelerated at TeV energies.
The $\gaga\to\gaga$ cross sections for $\mgg \geq$~5~GeV  are in the
hundreds fb,pb ranges for \pp,\pPb, and reach the $\mu$b level for \PbPb\ at the FCC, 
clearly showing the importance of the Z$^4$ enhancement of the photon fluxes in ion-ion collisions. The
number of LbyL events expected in ATLAS and CMS have been estimated with 
realistic $\gamma$ acceptance and efficiency cuts and integrated luminosities. In the \pp\ case, the dominant
background due to exclusive gluon-induced $\gaga$ production can be reduced imposing cuts on 
p$_{\rm T}^{\gaga}$ and pair acoplanarity, yet not to a level where the signal can be observed. The
signal/background ratio is better in the \pPb\ case but the small expected number of counts makes the
LbyL measurement challenging without (reachable) luminosity increases.  
Observation of the process is possible in \PbPb\ which provide
$\rm N_{\gaga}^{\rm excl}\approx$~70 elastic photon pairs per run after cuts at the LHC, with small 
backgrounds. The increase in $\gaga\to\gaga$ yields from LHC to FCC is of $\mathcal{O}(35)$ thanks to factors
of $\times$5 larger cross sections and luminosities, and $\times$2 in the experimental acceptance.
The measurement of elastic $\gaga$ scattering at the LHC will be the first-ever 
observation of such fundamental quantum mechanical process in the lab. At the FCC, 
the higher-masses of the produced diphoton system may be sensitive to new-physics effects
predicted in various SM extensions.\\

\noindent {\bf Acknowledgments -} We thank Lucian Harland-Lang for valuable discussions and for independent
cross-checks of some of our calculations.



\end{document}